\documentclass[draft]{agujournal2019}
\usepackage{url} 
\usepackage{lineno}
\usepackage[inline]{trackchanges} 
\usepackage{soul}
\usepackage{ragged2e}
\usepackage{amsmath}
\usepackage{tikz}
\usetikzlibrary{shapes,arrows.meta, positioning}

\justifying
\draftfalse

\journalname{Geophysical Research Letters}

\begin{document}

%
%


\title{Coupled Space Weathering: Nanophase Iron Formation by Micrometeoroid Impact and Solar Wind Sputtering}

%
%





\authors{Ziyu Huang\affil{1} and Masatoshi Hirabayashi\affil{1}}

\affiliation{1}{Daniel Guggenheim School of Aerospace Engineering, Georgia Institute of Technology, 620 Cherry
Street, Atlanta, GA 30332, USA.}







\correspondingauthor{Ziyu Huang}{zyuhuang@gatech.edu}




\begin{keypoints}
\item Micrometeoroid impacts significantly alter lunar surface binding energies and chemical compositions.
\item Monte Carlo simulation and atomistic simulation reveal distinct effects of impact and irradiation on lunar surface modifications.
\item Spatial variation in npFe$^0$ at impact sites reveals new targets for space weathering studies using remote sensing and sample analysis.
\end{keypoints}

%
%

%
%


\begin{abstract}

Understanding the interplay between micrometeoroid impacts and solar wind irradiation is crucial for interpreting lunar surface evolution. Using reactive molecular dynamics simulations and surface binding energy (SBE) analyses, this study investigates the coupled effects of these two dominant space weathering processes on lunar regolith composed of Fe$_2$SiO$_4$. Our simulations reveal that micrometeoroid impacts significantly modify the lunar surface, creating structurally heterogeneous zones with varying SBEs across {\color{black} micro}crater morphologies. Specifically, {\color{black} micro}crater floors exhibit enhanced surface cohesion due to high-density compaction, whereas {\color{black} micro}crater walls and ejecta show weakened structures. Applying Sigmund’s sputtering theory with these SBEs indicates differential sputtering yields for Fe, Si, and O, suggesting preferential retention of heavier elements like Fe. This selective sputtering mechanism supports the formation and growth of nanophase metallic iron (npFe$^0$) clusters, influencing the optical and compositional maturation of the lunar surface. These findings advance our understanding of lunar space weathering processes.

\end{abstract}

\section*{Plain Language Summary}

Micrometeoroid impacts and solar wind particles constantly alter the Moon’s surface. This study uses advanced computer simulations to see how these processes work together on a mineral called Fe$_2$SiO$_4$, found in lunar soil. The results show that impacts create different surface zones—some parts get compacted and strong, while others become weak and loose. These changes affect how atoms are knocked off the surface by solar wind, especially helping heavier atoms like iron stay behind. Over time, this helps form tiny iron particles that change how the Moon looks and behaves. Understanding this helps us prepare for future Moon missions.

%
%

%


%
%
%
%

\section{Introduction}

 Space weathering processes profoundly influence the surfaces of airless bodies such as the Moon, leading to alterations in their optical, chemical, and physical properties at various levels \cite{keller1993discovery,2000M&PS...35.1101P}. These processes have been reported to primarily result from two concurrent mechanisms: micrometeoroid impacts and solar wind irradiation. 
{\color{black}Micrometeoroid impacts (MMI) induce localized melting and vaporization, forming impact-melt glasses where nanophase metallic iron ($npFe^0$) is frequently found buried or embedded.} Solar wind irradiation implants energetic ions into the surface and sputter atoms from the surface. These processes are considered to drive the reduction of iron and promote the formation of nanophase metallic iron (npFe$^{0}$) particles \cite{Noble2007}. While both MMI and solar wind irradiation contribute to npFe$^{0}$ formation, solar wind irradiation primarily produces smaller npFe$^{0}$ particles, whereas MMI more effectively facilitates the clustering and aggregation of these particles within impact-melted regions \cite{2024NatAs...8.1110S}.
 These npFe$^{0}$ particles are responsible for the characteristic darkening and reddening of lunar regolith particles, with their optical effects varying based on their size and distribution.

While extensive studies have explored the individual effects of micrometeoroid impacts and solar wind irradiation on lunar regolith, the complex interplay and potential synergistic effects between these two primary space weathering agents remain incompletely understood. Recent analyses of lunar samples returned by the Chang'E-5 mission have revealed that larger iron particles are predominantly formed via high-energy micrometeoroid impacts, whereas smaller nanophase metallic iron (npFe$^{0}$) particles result mainly from continuous solar wind irradiation~\cite{Li2022, Guo2022}. These findings point to the existence of distinct formation pathways for iron nanoparticles on the Moon, highlighting the need to investigate how both processes operate in tandem or sequence under natural lunar conditions.

Despite these advances, significant knowledge gaps remain regarding the mechanisms by which micrometeoroid impacts and solar wind irradiation collectively influence regolith evolution over time. In particular, it is unclear how structural and compositional changes induced by impact events affect the subsequent interaction of the regolith with solar wind ions, and how this, in turn, governs the nucleation, mobility, and aggregation of iron nanoparticles at multiple scales. This lack of understanding limits the development of accurate models for the long-term evolution of lunar and other airless body surfaces, and poses challenges for interpreting compositional and spectral data from recent and forthcoming missions~\cite{Denevi2023, Chrbolkova2021, Guo2024}. Comprehensive studies addressing the coupled effects of these weathering processes are therefore crucial for advancing planetary surface science and unraveling the history of solar system materials.

This study employs advanced computational simulations—combining molecular dynamics with theoretical sputtering yield calculations—to systematically investigate the coupled effects of micrometeoroid hypervelocity impacts and solar wind irradiation on lunar regolith. {\color{black} Throughout this study, 'crater' is used to denote these micro-scale impact morphologies.} By elucidating the complex interplay between impact-induced vaporization, shock-driven structural modification, and subsequent energetic particle sputtering, we reveal how these synergistic processes drive the evolution of regolith chemistry, promote nanophase iron (npFe$^{0}$) formation, and control volatile retention. Our approach enables in-depth analysis of mechanisms governing the formation, redistribution, and aggregation of npFe$^{0}$ particles, as well as the depth-dependent alteration of the regolith structure. These new insights allow for more accurate interpretation of mission data from Artemis~III and Chang’E~6—including observed differences in space weathering signatures between the lunar nearside and farside~\cite{zhang2025space}—and directly inform the development of durable materials, infrastructure, and scientific instruments for long-term lunar exploration. Ultimately, this research advances our understanding of both the historical and ongoing space weathering processes that shape airless planetary surfaces, with broad implications for planetary science and future deep space missions.

\section{Method}

\subsection{Impact Simulation by Reactive Molecular Dynamics}

Molecular dynamics (MD) simulations are performed using Large-scale Atomic Molecular Massively Parallel Simulator (LAMMPS) to simulate the micrometeoroid impacts on the lunar surface. The initial conditions are set to replicate a typical impact event by defining the speed, angle, and mass of the impacting micrometeoroid. The simulation parameters are derived from \cite{cintala1992impact,pokorny2019meteoroids}. 
As an Fe-rich silicate mineral commonly observed in lunar regolith, fayalite (Fe$_2$SiO$_4$) was selected as a representative phase \cite{Papike1998}.
The simulation domain is set to be periodic in all three dimensions, allowing for the use of a smaller volume to represent the general behaviors in a larger element in all three spatial directions.


  
Reactive molecular dynamics (RMD) explicitly accounts for bond formation and breaking between atoms, enabling the simulation to capture chemical reactions dynamically during the trajectory. The same method have been used to simulate micrometeoroid impact induced lunar volatile formation \cite{huang2021molecular}. The interactions between atoms are described using a reactive force field (ReaxFF) potential \cite{van2001reaxff}, which computes the total energy as a sum of multiple bonded and non-bonded contributions:

\begin{equation}
\begin{aligned}
E_{\text{ReaxFF}}(\{\vec{r}_{ij}\}, & \{\vec{r}_{ijk}\}, \{\vec{r}_{ijkl}\}, \{q_{i}\}, \{BO_{ij}\}) = E_{\text{bond}} + E_{\text{lp}} + E_{\text{over}} + E_{\text{under}} + \\
& E_{\text{val}} + E_{\text{pen}} + E_{\text{coa}} + E_{\text{tors}} + E_{\text{conj}} + E_{\text{hbond}} + E_{\text{vdWaals}} + E_{\text{Coulomb}}
\end{aligned}
\end{equation}


\noindent The total energy $E_{ReaxFF}$ depends on the interatomic distances between atomic pairs, $\vec{r}_{ij}$, triplets, $\vec{r}_{ijk}$, and quadruplets, $\vec{r}_{ijkl}$, as well as on atomic charges $q_i$ and the bond orders $BO_{ij}$ of atomic pairs. The valence interactions consist of bonding energy ($E_{bond}$), lone-pair energy ($E_{lp}$), overcoordination energy ($E_{over}$), undercoordination energy ($E_{under}$), valence-angle energy ($E_{val}$), penalty energy ($E_{pen}$), three-body conjugation energy ($E_{coa}$), torsion-angle energy ($E_{tors}$), four-body conjugation energy ($E_{conj}$), and hydrogen bonding energy ($E_{hbond}$). Noncovalent interactions include the van der Waals energy ($E_{vdWaals}$) and Coulomb energy ($E_{Coulomb}$), both of which are modulated by a taper function.

\subsection{Surface Binding Energy {\color{black} Calculation}}

To quantify the surface binding energy (SBE) at micrometeoroid impact craters and irradiated sites, we adopt an iterative computational approach similar to the method described by \citeA{morrissey2022simulating} and \citeA{yang2014atomic}. For each substrate, {\color{black} maintained at a constant temperature of $300\text{ K}$}, a randomly selected surface atom (e.g., Fe, O  or Si) was assigned a specific kinetic energy {\color{black}along the local surface normal}, and its trajectory was tracked over time to monitor its position and residual energy. The minimum kinetic energy required to completely remove the atom from the surface---defined as the point at which the atom is sufficiently far away that it no longer experiences any attractive or repulsive forces and its energy remains constant---was determined iteratively. 

As the SBE {\color{black} calculation} requires extensive iterative NVE simulations, we limited our calculation to the impact region, which is spatially confined within a hemispherical volume of radius 70~\AA{} centered at the impact site. This localized approach not only focuses on the most relevant area affected by the micrometeoroid impact but also dramatically reduces the computational load. In total, 35,846 atoms were included for SBE evaluation—substantially fewer than the full system used for the micrometeoroid impact simulations (318,459 atoms). This reduction enabled efficient sampling of a statistically meaningful set of surface atoms, while preserving the essential local structural and chemical features of the impact-modified region. Each SBE determination was repeated for selected Fe, Si, and O atoms distributed throughout the exposed {\color{black} micro}crater surface and ejecta rim, ensuring that the calculated {\color{black}surface} binding energies reflect the heterogeneity introduced by impact-induced disorder and compositional gradients.

\subsection{Sputtering Yield from TRIM Simulation}
{\color{black}
To investigate the effects of solar wind irradiation on the impacted lunar surface, we performed simulations using the SRIM (Stopping and Range of Ions in Matter)/TRIM (TRansport of Ions in Matter) code \cite{hofsass2014simulation,biersack1980monte}. This well-established Monte Carlo simulation tool models the stopping and range of ions in matter, providing detailed statistics on sputtering and ion implantation. This direct simulation method provides a significant advantage over analytical approaches like Sigmund's sputtering theory \cite{Sigmund1969}. While later theoretical models based on \citeA{Sigmund1969}'s model such as  \citeA{yamamura1996energy} provide a valuable first-order approximation and empirical equations, they are typically idealized for mono-elemental targets. In contrast, the TRIM simulation inherently models {\color{black}collision cascade using the Binary Collision Approximation (BCA)}, naturally accounting for the distinct masses and ejection probabilities of Fe, Si, and O atoms within the same collisional cascade. This provides a more detailed and physically robust prediction of preferential sputtering from a complex mineral, a critical factor for accurately modeling the compositional evolution of the lunar surface.

We simulated a beam of 1 keV H$^+$ ions which represent the typical energy and composition of the solar wind, onto the substrate. Although the micrometeoroid impact site is largely amorphous, we assumed the bulk material composition to be fayalite (\(\text{Fe}_2\text{SiO}_4\)) for the purpose of calculating the stopping power and simulating the energy cascade process within TRIM. The simulation calculates the sputtering yield, defined as the average number of target atoms ejected per incident ion, by tracking the collision cascades initiated by each ion. The trajectories of 100000 of individual ions are simulated as they penetrate the target material, transferring energy to the target atoms through a series of binary collisions. This process accounts for both nuclear and electronic stopping energy losses. To assess the influence of near-surface bonding on material ejection, the simulation was run using four different cases for the surface binding energy (SBE). From each of these four runs, we gathered the respective sputtering yields for each atomic species (Fe, Si, and O). 
}

\section{Results and Discussions}

\subsection{Micrometeoroid Impact Setup and Surface Amorphization}



The impact setup 
uses a 4 nm projectile striking a 40 nm Fe$_2$SiO$_4$ half-hemisphere substrate at 12 km/s and a 45° incidence angle. The choice of these parameters is informed by prior studies: the 45° angle is considered the most probable micrometeoroid impact angle on planetary surfaces \cite{melosh1997impact}, while the 12 km/s velocity is identified as the typical encounter speed of micrometeoroids on the lunar surface\cite{cintala1992impact,pokorny2019meteoroids}. In the simulation, the entire impact process unfolds over roughly 50 femtoseconds, during which the projectile transfers its energy to the target, generating ejecta that rapidly leave the simulation box. The system is carefully monitored to ensure that no ejected atoms fall back onto the surface, preventing artificial re-impact effects.

\begin{figure}
\centering
\includegraphics[width=\linewidth]{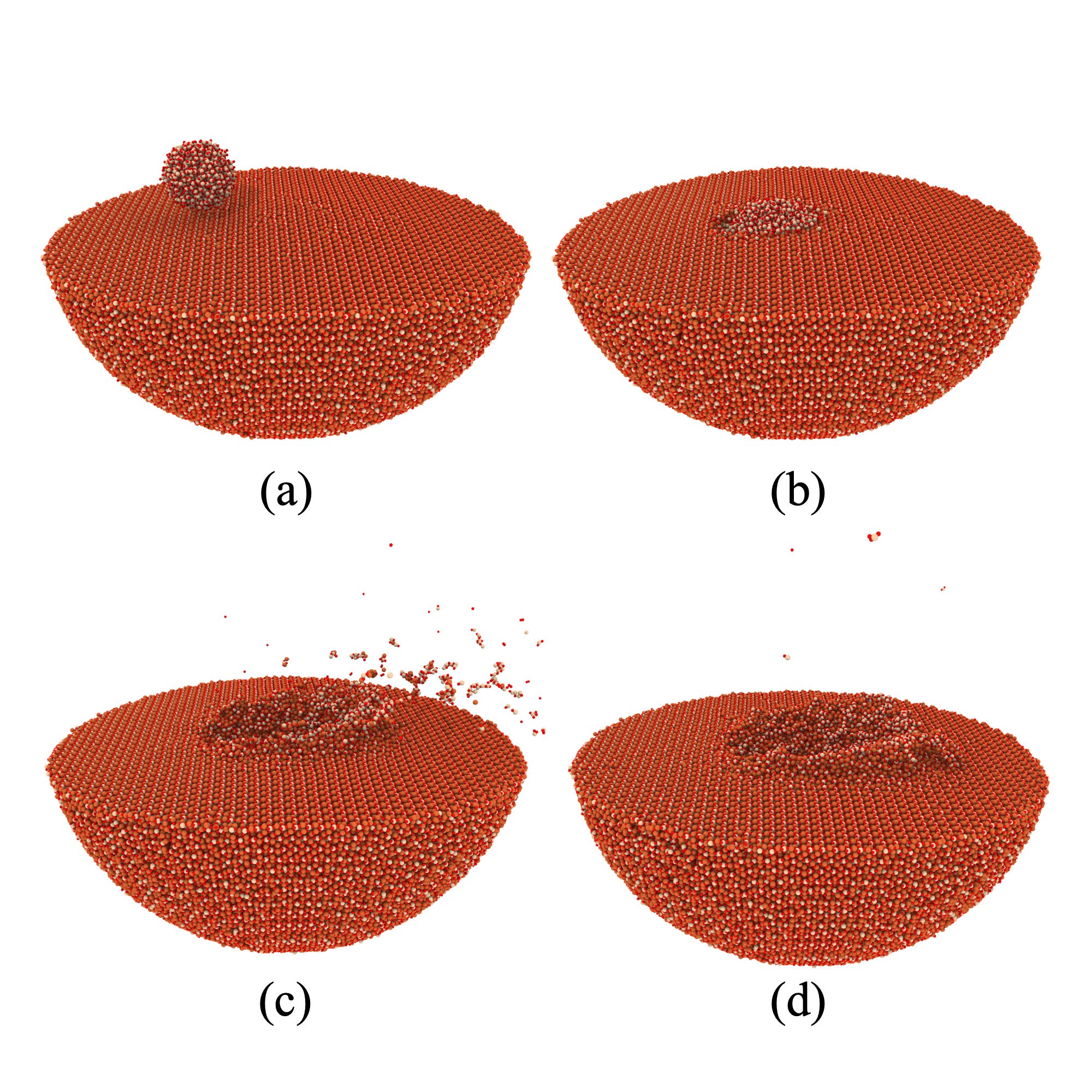}
\caption{
Schematic illustration of a micrometeoroid impact sequence: (a) Initial setup showing a 4 nm micrometeoroid approaching a pristine $Fe_2SiO_4$ mineral surface; (b) Moment of contact between the micrometeoroid and the surface; (c) Formation of ejecta as atomic structures are disrupted; (d) Post-impact state depicting the amorphized $Fe_2SiO_4$ surface. {\color{black} Atom colors represent Oxygen (red), Iron (orange), and Silicon (tan).}
}
\label{fig:impact_schematic}
\end{figure}

As shown in Figure~\ref{fig:impact_schematic}, the high-velocity impact generates shock waves that propagate outward from the contact point, producing pressures exceeding tens of GPa and transient temperatures that can reach over 4000 K. These extreme conditions cause rapid, localized modification of the target’s crystalline lattice—breaking down its ordered atomic structure and generating a variety of defects such as vacancies, dislocations, and grain boundary distortions. The combination of intense compression and thermal excitation leads to the swift loss of long-range order, resulting in the formation of amorphous zones interspersed with residual crystalline domains.



\subsection{Surface Binding Energy across {\color{black}Microcrater} Morphologies}

\begin{figure}[!h]
\includegraphics[width=\linewidth]{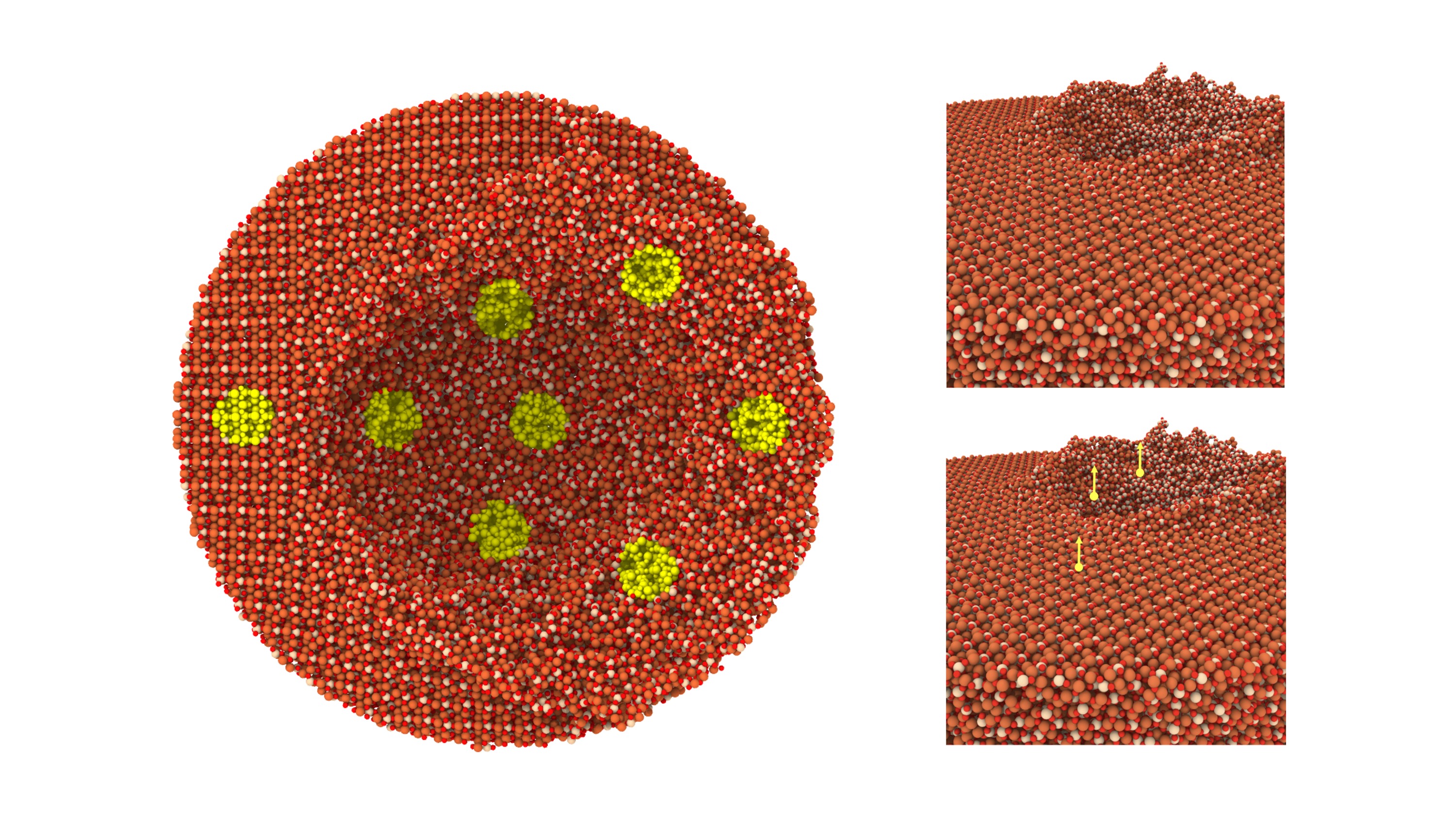}
\caption{Left panel: Sampling locations for surface binding energy {\color{black} calculation}s, with selected atoms colored in yellow. Locations include the pristine surface (undisturbed crystalline region), crater wall (compressed and sheared zone), crater floor (heavily amorphized region), and ejecta zone (displaced and re-deposited material). Upper right panel: Close-up view of the amorphous impact site on the regolith testbed surface; Lower right panel: illustration of atoms being artificially given kinetic energy until they detach, demonstrating the procedure used to measure surface binding energy
}
\label{fig:Figure2}
\end{figure}

Figure~\ref{fig:Figure2} provides a close-up view of the amorphous surface layer generated at the impact site, highlighting the disrupted atomic arrangements left behind after the collision. In this analysis, atoms located on the surface are artificially assigned progressively increasing kinetic energy to simulate the detachment process, a method designed to probe the energy required to overcome local binding forces. As visualized by the three arrows in the figure, this procedure tracks the precise moment when individual atoms escape from the surface, allowing direct {\color{black} calculation} of the surface binding energy.

{\color{black}The left panel of Figure~\ref{fig:Figure2}} highlights the specific surface regions selected for {\color{black}surface} binding energy {\color{black} calculation}s, including pristine areas, the crater wall, the crater floor, and the ejecta zone. The pristine surface is sampled to establish a reference for the unaltered, crystalline state of the material prior to impact. The crater wall is chosen because it experiences both compression and shearing during the impact, producing mixed crystalline-amorphous structures. The crater floor is targeted as the region most heavily compacted and amorphized under the direct force of the projectile, representing the zone of maximum structural disruption. Finally, the ejecta zone encompasses material displaced and redeposited by the impact, including particles that have undergone both mechanical ejection and partial melting or vaporization. These distinct locations are selected to reflect the diverse physical conditions produced across the crater morphology, allowing for detailed site-specific {\color{black} calculation}s.


Surface binding energy (SBE) calculations at the selected sites (Table 2) show distinct differences tied to local structural conditions. On the pristine surface, values are 4.3 eV (Fe), 4.7 eV (Si), and 2.0 eV (O), reflecting the stable, ordered lattice. At the crater wall, SBE drops to 2.7 eV (Fe), 2.8 eV (Si), and 1.5 eV (O). {\color{black}This reduction is quantitatively linked to a decrease in the average coordination number (CN) of surface atoms (Table S1), where impact-induced amorphization leads to an under-coordinated state with fewer immediate neighbors to maintain chemical bonds.} The crater floor shows the highest readings—5.8 eV (Fe), 6.8 eV (Si), and 3.4 eV (O)—{\color{black}consistent with an observed ~20\% increase in local coordination numbers (e.g., CN of iron atoms on the surface increases to ~4.25 compared to 3.94 for pristine surface)} formed under extreme compressive stress. In contrast, the projectile material yields the lowest energies ($2.0 \text{ eV for Fe}$), a result of its fragmented network following the collision {\color{black} with CN of iron atoms drop to 2.14}.  {\color{black}This is primarily due to the severe amorphization of the projectile material, which results in a significantly lower coordination number for the constituent atoms. The transition from a structured lattice to a disordered, under-coordinated state means that each atom has fewer immediate neighbors to maintain chemical bonds, directly leading to the observed reduction in surface binding energy.}
Unlike the crater floor, which {\color{black}experiences} compressive forces upon impact, the ejecta material is expelled outward with greater freedom to expand. This rapid expansion facilitates bond breaking and structural disruption, resulting in a more disordered, energetically relaxed state. These site-specific energetic differences highlight how impact dynamics, including local confinement and expansion, shape the post-impact energy landscape across cratered surfaces.

\begin{table}[!h]
\centering
\begin{tabular}{lcccc}
\hline
\textbf{Element} & \textbf{Pristine Surface } & \textbf{Crater Wall} & \textbf{Crater Floor} & \textbf{Ejecta} \\
\hline
Fe & 4.3 & 2.7 & 5.8 & 2.0 \\
Si & 4.7 & 2.8 & 6.8 & 1.4 \\
O  & 2.0 & 1.5 & 3.4 & 1.0 \\
\hline
\end{tabular}

\caption{Average Surface binding energies ($U_s$) used for sputtering yield calculations under different surface conditions (Energy unit in eV)}
\label{tab:SBE_values}
\end{table}

{\color{black}
\subsection{Sputtering Yield}





The calculated sputtering yields (Y) for iron (Fe), silicon (Si), and oxygen (O) across four distinct surface terrains—pristine, crater wall, crater floor, and ejecta—are presented in Table \ref{tab:Sputtering_Yields}.  These values, given in units of atoms sputtered per incident ion, were derived using TRIM calculations with surface binding energies obtained from our {\color{black} calculation}s in Table \ref{tab:SBE_values}. The table also includes the total sputtering yield (Y$_{Total}$) and the relative sputtering yield of iron (Y$_{Fe}$/Y$_{Total}$) for each location.

\begin{table}[!h]
\centering
\begin{tabular}{ccccc}
\hline
\textbf{Element}     & \textbf{Pristine} & \textbf{Crater Wall} & \textbf{Crater Floor} & \textbf{Ejecta} \\ \hline
Y$_{Fe}$             & 0.00353           & 0.00554              & 0.00288               & 0.00733         \\
Y$_{Si}$             & 0.00202           & 0.00402              & 0.00175               & 0.00723         \\
Y$_{O}$              & 0.01330           & 0.02310              & 0.01130               & 0.03360         \\
Y$_{Total}$          & 0.01890           & 0.03270              & 0.01590               & 0.04820         \\
Y$_{Fe}$/Y$_{Total}$ & 0.18677           & 0.16942              & 0.18113               & 0.15207         \\ \hline
\end{tabular}
\caption{Sputtering yields of Fe, O, and Si atoms at different surface locations following micrometeoroid impact.}
\label{tab:Sputtering_Yields}
\end{table}

A primary observation from the data is the significant variation in the total sputtering yield across the different terrains. The crater floor exhibits the lowest total yield with $Y_{\text{total}} = 0.01590$, while the ejecta material shows the highest at $Y_{\text{total}} = 0.04820$. This trend is directly linked to the physical state of the surface material. The crater floor, having been subjected to immense pressure during the impact event, is expected to be more compact and possess a higher average surface binding energy, which suppresses the release of atoms. In contrast, the ejecta consists of loosely bonded, {\color{black}low-density, under-coordinated system} with a lower effective surface binding energy, making it substantially more susceptible to sputtering. The crater wall and pristine surfaces show intermediate total yields of 0.03270 and 0.01890, respectively.

To assess the elemental fractionation and subsequent surface enrichment, we calculated the relative sputtering yield of iron ($Y_{\text{Fe}}/Y_{\text{total}}$). This ratio is a crucial metric, as it is inversely proportional to the surface accumulation rate of Fe due to sputtering yield; a lower value indicates that Fe is preferentially retained on the surface while other elements are removed. The benchmark for this fractionation is the original stoichiometric composition of the target material, {\color{black}fayalite} ($\text{Fe}_2\text{SiO}_4$). In this mineral, the atomic fraction of iron is 2 out of 7 total atoms (2 Fe, 1 Si, 4 O), or approximately 0.286. As shown in Table~\ref{tab:Sputtering_Yields}, the calculated $Y_{\text{Fe}}/Y_{\text{total}}$ values for all four terrains (ranging from 0.15207 to 0.18677) are well below this stoichiometric threshold. This confirms that solar wind sputtering preferentially removes the lighter species (O and Si) over the heavier Fe in all cases, driving the enrichment of iron on the surface, which is the foundational process for forming nanophase metallic iron (npFe$^0$).

Furthermore, the data reveal that impact-induced material alteration significantly enhances this elemental differentiation. The pristine surface has the highest relative Fe sputtering yield at 0.18677, whereas all impact-affected regions---the crater wall (0.16942), crater floor (0.18113), and ejecta (0.15207)---exhibit lower values. This suggests that the amorphization and structural disordering caused by the micrometeoroid impact weaken the bonds of lighter elements more effectively than those of the heavier iron, thereby amplifying the preferential sputtering effect. The impact process, therefore, not only redistributes material but also chemically sensitizes it to more rapid and pronounced space weathering.

Among the three different crater locations, the results strongly indicate that the ejecta region is the most favorable site for $npFe^0$ formation. The ejecta region pairs the highest overall sputtering rate ($Y_{\text{total}} = 0.04820$) with the most efficient iron retention (the lowest $Y_{\text{Fe}}/Y_{\text{total}}$ of 0.15207). {\color{black}These distinct yield ratios represent a systematic chemical divergence that is critical for understanding regolith evolution over geological timescales. A ~25\% relative reduction in iron loss in the ejecta compared to the crater floor constitutes a primary mechanism for differential iron enrichment. When integrated over the long-term exposure age of the lunar surface, this sustained variance in sputtering efficiency governs the cumulative mass balance of the soil. This leads to an accelerated buildup of an Fe-rich surface layer in specific regions, directly dictating the localized nucleation and growth rates of $npFe^0$ particles and ultimately shaping the spatial heterogeneity of lunar optical maturity.}

\section{Conclusion}

Our results provide new insights into the coupled effects of micrometeoroid impacts and solar wind irradiation on lunar surface evolution. Apollo sample studies together with recent analyses of both Chang’E-5 and Chang’E-6 samples have underscored the dominant role of solar wind weathering on the lunar farside and revealed potentially distinct weathering pathways compared to nearside terrains, particularly regarding the mechanisms of nanophase metallic iron (npFe$^0$) formation~\cite{Noble2007,gopon2019complementary,taylor2001lunar,2022GeoRL..4997323G,2022NatAs...6.1156L,2022NSRev...9B.188L,2024NatAs...8.1110S,2023NatAs...7..280X}. Our molecular dynamics simulations of Fe$_2$SiO$_4$ impacts, combined with measured surface binding energy (SBE) variations and Sigmund theory sputtering yields, offer a mechanistic explanation for how impact-induced heterogeneity—such as the creation of dense, compacted crater floors and weakened ejecta surfaces—can set the stage for differential npFe$^0$ formation under prolonged solar wind exposure. 
Notably, this work demonstrates that the distribution of npFe$^0$ varies significantly across different locations within an impact crater, reflecting the diverse temperature and pressure conditions experienced during the impact process. This spatial variation in npFe$^0$ abundance and properties presents a critical opportunity for future remote sensing observations and targeted sample analyses, which can directly test model predictions by comparing npFe$^0$ distributions and spectral changes across micrometeoroid impact sites. While our current model primarily addresses structural modification and sputtering, further work should explore whether altered zones facilitate disproportionation reactions or promote npFe$^0$ coarsening, particularly where surfaces have been preconditioned by impacts.

{\color{black}For a single microcrater, it should be noted that this localized Fe enrichment is inherently temporary due to the significant disparity between weathering timescales. Stochastic impact gardening can re-mix the regolith and overturn the uppermost layer, this process requires orders of magnitude more time than the initial modification by solar wind irradiation \cite{costello2018mixing,costello2020impact}, which occurs over years to decades under a constant flux ($\sim 10^8 \text{ cm}^{-2}\text{s}^{-1}$). Initially, varying surface binding energies across the microcrater drive the preferential ejection of non-Fe species. As surface Fe accumulates, the sputtered Fe fraction ($Y_{Fe}/Y_{total}$) rises until the irradiated layer reaches a steady state of stoichiometric sputtering \cite{behrisch2007sputtering}, capping further Fe concentration. Over these longer timescales, prolonged solar wind irradiation will ultimately erode the microcrater topology entirely, locally erasing the spatial heterogeneity. However, because the timescale for this rapid initial Fe enrichment is substantially shorter than that of complete microcrater removal, the constant flux of new micrometeoroid impacts ensures that these temporary, spatially heterogeneous, Fe-enriched microenvironments are continuously regenerated at different microcraters and dynamically maintained across the global lunar surface.}

Beyond lunar science, our findings highlight a broadly applicable weathering mechanism by impact-driven surface modification followed by differential sputtering, which likely shapes the surfaces of many airless bodies, including Mercury, asteroids, and outer planet moons. The extent and manifestation of this process depend strongly on local factors such as micrometeoroid flux, local plasma environment, and especially surface mineralogy, as the Fe content and diversity of mineral species directly influence both impact physics and sputtering responses. For instance, Mercury experiences higher micrometeoroid impact velocities that enhance both amorphization and sputtering \cite{cintala1992impact}, and Galilean moons are primarily modified by energetic magnetospheric ions. These distinct environmental factors shape a diverse range of impact-driven amorphization and sputtering outcomes across airless bodies in the solar system. In addtion, although our study advances mechanistic understanding using Fe$_2$SiO$_4$ as Fe-bearing mineral species, real planetary surfaces exhibit a wider range of mineralogical diversity and Fe proportions that can further modulate weathering pathways. {\color{black}While SRIM/TRIM serves as a widely utilized benchmark for evaluating the first-order sensitivity of sputtering yields to SBE variations, it is important to acknowledge that absolute yield values can exhibit sensitivity to the choice of simulation framework. {\color{black} Recent comparative analyses \cite{brotzner2025solar} have explored how different binary collision approximation (BCA) models, such as SDTrimSP and TRIM, differ in their predicted sputtering yields.} While the relative differences and the resulting $npFe^0$ distribution patterns are driven by the MD-calculated SBE and remain robust, future efforts will involve the use of SDTrimSP to provide more precise absolute yield estimations. Combining these refined yields with regional gardening rates will allow for a more accurate assessment of the geological timescales required for the development of mature regolith.} Additional work integrating diverse mineral targets, longer simulation timescales, and experimental validation will enable more accurate and comprehensive modeling of space weathering processes across the solar system.

\acknowledgments

This research was supported in part through research cyberinfrastructure resources and services provided by the Partnership for an Advanced Computing Environment (PACE) at the Georgia Institute of Technology, Atlanta, Georgia, USA. Z.H. and M.H. are supported by SSERVI-CLEVER (NNH22ZDA020C/80NSSC23M022). M.H. acknowledges support by VIPER (80NSSC24K0682) and SSERVI/RASSLE (80NSSC24M0016).

\section*{Conflict of Interest Statement}

The authors have no conflicts of interest to disclose.

\section*{Open Research Section}
The data that support the findings of this study have been publicly archived and are accessible through Zenodo \cite{huang_2026_18246651}.

\bibliography{Impact_Sputtering,spaceweathering}

\end{document}